\documentclass{elsart}
\usepackage{epsfig}
\usepackage{bm}% bold math
\begin{document}
\begin{frontmatter}
\title{\boldmath Partial wave analysis of $J/\psi \to \gamma \phi \phi$}
\begin{small}
\begin{center}

M.~Ablikim$^{1}$,              J.~Z.~Bai$^{1}$,   Y.~Bai$^{1}$,
Y.~Ban$^{11}$, X.~Cai$^{1}$,                  H.~F.~Chen$^{16}$,
H.~S.~Chen$^{1}$,              H.~X.~Chen$^{1}$, J.~C.~Chen$^{1}$,
Jin~Chen$^{1}$
% ...
\\
\vspace{0.2cm}
(BES Collaboration)\\
\vspace{0.2cm} {\it
$^{1}$ Institute of High Energy Physics, Beijing 100049, People's Republic of China\\
% ...
$^{11}$ Peking University, Beijing 100871, People's Republic of China\\
% ...
$^{16}$ University of Science and Technology of China, Hefei 230026,
People's Republic of China\\
% ...
}
\end{center}

\vspace{0.4cm}
\end{small}
\date{\today}

\maketitle
\normalsize
\begin{abstract}
Using $5.8 \times 10^7 J/\psi$ events collected in the BESII
detector, the radiative decay $J/\psi \to \gamma \phi
\phi \to \gamma K^+ K^- K^0_S K^0_L$ is studied. The
$\phi\phi$ invariant mass distribution exhibits a near-threshold
enhancement that peaks around 2.24~GeV/$c^{2}$.
%An analysis of
%angular correlations between the decay particles indicates that the
%structure is predominantly even spin and odd parity.
A partial wave analysis shows that the structure is dominated by a
$0^{-+}$ state ($\eta(2225)$) with a mass of
$2.24^{+0.03}_{-0.02}{}^{+0.03}_{-0.02}$~GeV/$c^{2}$ and a width of
$0.19 \pm 0.03^{+0.06}_{-0.04}$~GeV/$c^{2}$. The product branching
fraction is: $Br(J/\psi \to \gamma \eta(2225))\cdot Br(\eta(2225)\to
\phi\phi)
 = (4.4 \pm 0.4 \pm 0.8)\times 10^{-4}$.
\end{abstract}
\end{frontmatter}
\section{Introduction}   \label{introduction}
In addition to the spectrum of ordinary light $q\bar{q}$ meson
states, QCD-motivated models predict a rich spectrum of $gg$
glueballs, $q\bar{q}g$ hybrids and $qq\bar{q}\bar{q}$ four quark
states~\cite{nonqq}. Radiative $J/\psi$ decays provide an excellent
laboratory for testing these predictions, and systems of two vector
particles have been intensively examined for signatures of gluonic
bound states~\cite{glueball}. Pseudoscalar enhancements in
$\rho\rho$ and $\omega\omega$ final states have been seen in
radiative $J/\psi$ decays~\cite{rad1,rad2,rad3,rad4, BES-WW}.
Recently, a near-threshold scalar, the $f_0(1810)$ or $X(1810)$ was
reported in the $\omega\phi$ invariant mass distribution from the
doubly OZI suppressed decays of
$J/\psi\to\gamma\omega\phi$~\cite{x1810}, thereby adding an
additional puzzle to the already confusing spectrum of low-lying
scalar mesons~\cite{bugg,scalar}.
%%You should reference some review (or something) that
%%discusses the scalar meson puzzle

Structures in the $\phi\phi$ invariant-mass spectrum have been
observed by several experiments both in the reaction $\pi^{-}p\to\phi\phi
n$~\cite{pip} and in radiative $J/\psi$ decays~\cite{mk3,dm21,dm22}.
%A pseudoscalar signal is observed in the low region of $\phi\phi$
%invariant mass spectrum, known as $\eta(2225)$ in Particle Data
%Group (PDG) Tables~\cite{pdg2006} with a mass of $M=
%2220\pm18$~MeV/$c^2$ and a width of $\Gamma=
%150^{+300}_{-60}\pm60$~MeV/$c^2$
The $\eta(2225)$ was first observed by the MARK III collaboration in
$J/\psi$ radiative decays $J/\psi \to \gamma\phi\phi$~\cite{mk3}. A
fit to the $\phi\phi$ invariant-mass spectrum gave a mass of $2230
\pm25\pm15$~MeV/$c^2$ and a width of
$150^{+300}_{-60}\pm60$~MeV/$c^2$. The production branching
fractions are $Br(J/\psi \to \gamma \eta(2225))\cdot
Br(\eta(2225)\to \phi\phi)
 = (3.3 \pm 0.8 \pm 0.5)\times 10^{-4}$ for the $\gamma K^+ K^- K^+ K^-$ mode and $Br(J/\psi \to \gamma \eta(2225))\cdot
Br(\eta(2225)\to \phi\phi)
 = (2.7 \pm 0.6 \pm 0.6)\times 10^{-4}$ for the $\gamma K^+ K^- K^0_S K^0_L$ mode. An angular analysis of the structure found
it to be consistent with a $0^{-+}$ assignment. It was subsequently
observed by the DM2 collaboration, also in $J/\psi \to \gamma \phi
\phi$ decays~\cite{dm21,dm22}.

In this letter we present results from a high statistics study of
$J/\psi \to \gamma \phi \phi$ in the $\gamma K^+ K^- K^0_S K^0_L$
final state, using a sample of $5.8 \times 10^7 J/\psi$ events
collected with the BESII detector at the Beijing Electron-Positron
Collider (BEPC). The presence of a signal around 2.24~GeV/$c^2$ and
its pseudoscalar character are confirmed, and the mass, width, and
branching fraction are determined by a partial wave analysis (PWA).

\section{ \boldmath BES detector and Monte Carlo simulation}
BESII is a large solid-angle magnetic spectrometer that is described
in detail in Ref.~\cite{BESII}. Charged particle momenta are
determined with a resolution of $\sigma_p/p = 1.78 \% \sqrt{1+p^2}$
(with $p$ in GeV/$c$) in a 40-layer cylindrical main drift chamber
(MDC).  Particle identification is accomplished using specific
ionization ($dE/dx$) measurements in the MDC and time-of-flight
(TOF) measurements in a barrel-like array of 48 scintillation
counters. The $dE/dx$ resolution is $\sigma_{dE/dx}$ = 8.0\%;  the
TOF resolution is $\sigma_{TOF}$ = 180 ps for the Bhabha events.
Outside of the TOF counters is a 12-radiation-length barrel shower
counter (BSC) comprised of gas tubes interleaved with lead sheets.
The BSC measures the energies and directions of photons with
resolutions of $\sigma_E/E \simeq 21 \% /\sqrt{E}$ (with $E$ in
GeV), $\sigma_{\phi}$ = 7.9  mrad, and $\sigma_z$ = 2.3 cm.
The iron flux return of the magnet is instrumented with three double
layers of counters that are used to identify muons.

In this analysis, a GEANT3-based Monte Carlo (MC) simulation program
(SIMBES)~\cite{SIMBES}, which includes detailed consideration of
the actual detector responses (such as dead electronic channels),
is used. The
consistency between data and Monte Carlo has been checked
with many high-purity physics channels~\cite{SIMBES}.

\section{\boldmath Event selection}
Since the $K^0_L$ is difficult to identify in BESII, its detection
is not required in the selection of $J/\psi\to\gamma
K^+K^-K^0_SK^0_L$ events. $J/\psi\to\gamma
K^+K^-\pi^+\pi^-(K^0_L)_{miss}$ candidates are selected from events
with four charged tracks with net charge zero in
the MDC and with one or two isolated photons in the BSC.
%\subsection{\boldmath Charged particle identification}
Charged tracks are required
to be well fitted to a helix, be within the polar angle region
$|\cos\theta| < 0.8$, and have a transverse momentum larger than
50~MeV/$c$.
%, and have the point of closest approach of the track to the
%beam axis within 2 cm of the beam axis and within 20 cm from the
%center of the interaction region along the bean line.
For each track, the time-of-flight (TOF) and specific ionization
($dE/dx$) measurements in the MDC are combined to form a particle
identification confidence level for the $\pi, K$ and $p$ hypotheses; the
particle type with the highest confidence level is assigned to each
track. The four selected charged tracks are required to consist of
an unambiguously identified $K^+$, $K^-$, $\pi^+$ and $\pi^-$
combination.
%\subsection{\boldmath Photon identification}
Each candidate photon is required to have an energy deposit in the
BSC greater than 60~MeV, to be isolated from charged tracks by more
than $20^\circ$,
%to come from the interaction region,
%to be with the angle between the development direction of the
%cluster in BSC and the direction from the interaction point to
%the first hit layer of the BSC  less than 30$^\circ$,
have an angle between the cluster development direction in the BSC
and the photon emission direction less than  30$^\circ$, and have its
first hit in the BSC within the first six radiation lengths.
%\subsection{\boldmath Event selection criteria}
%Events are required to have four charged tracks with net charge zero
%and either one or two photon candidates.
A one-constraint(1C)
kinematic fit is performed with the $J/\psi \to \gamma
K^+K^-\pi^+\pi^-{(K^0_L)}_{miss}$ hypothesis.  For events with two isolated
photons,  the one that gives  the smallest $\chi^{2}({\gamma
K^+K^-\pi^+\pi^-{(K^0_L)}_{miss}})$ is selected. In order to improve
the $K^0_SK^0_L$ mass resolution, a 2C-kinematic fit is performed by
adding a $K^0_S$ mass constraint. Events with
$\chi^{2}_{1C}({\gamma K^+K^-\pi^+\pi^-{(K^0_L)}_{miss}})<20$ and
$\chi^{2}_{2C}({\gamma K^+K^-K^0_S{(K^0_L)}_{miss}})<20$ are
retained.

\begin{figure}[htbp]
%   \vskip -1cm
      \centerline{
                       \psfig{file=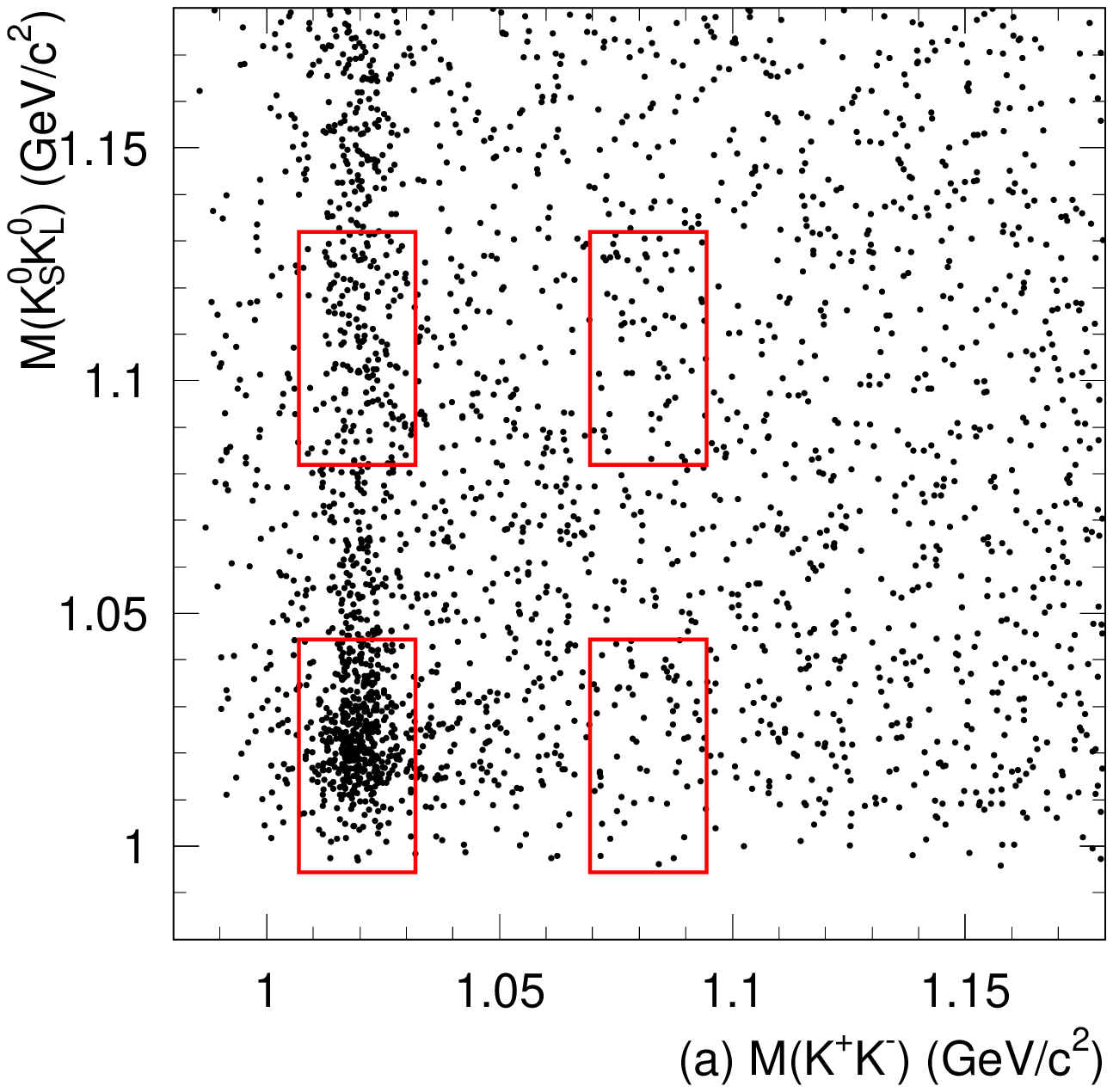,width=7.5cm,height= 7.5cm,angle=0}
 %                   \put(-40,100){(a)}
 }
   \centerline{
   \psfig{file=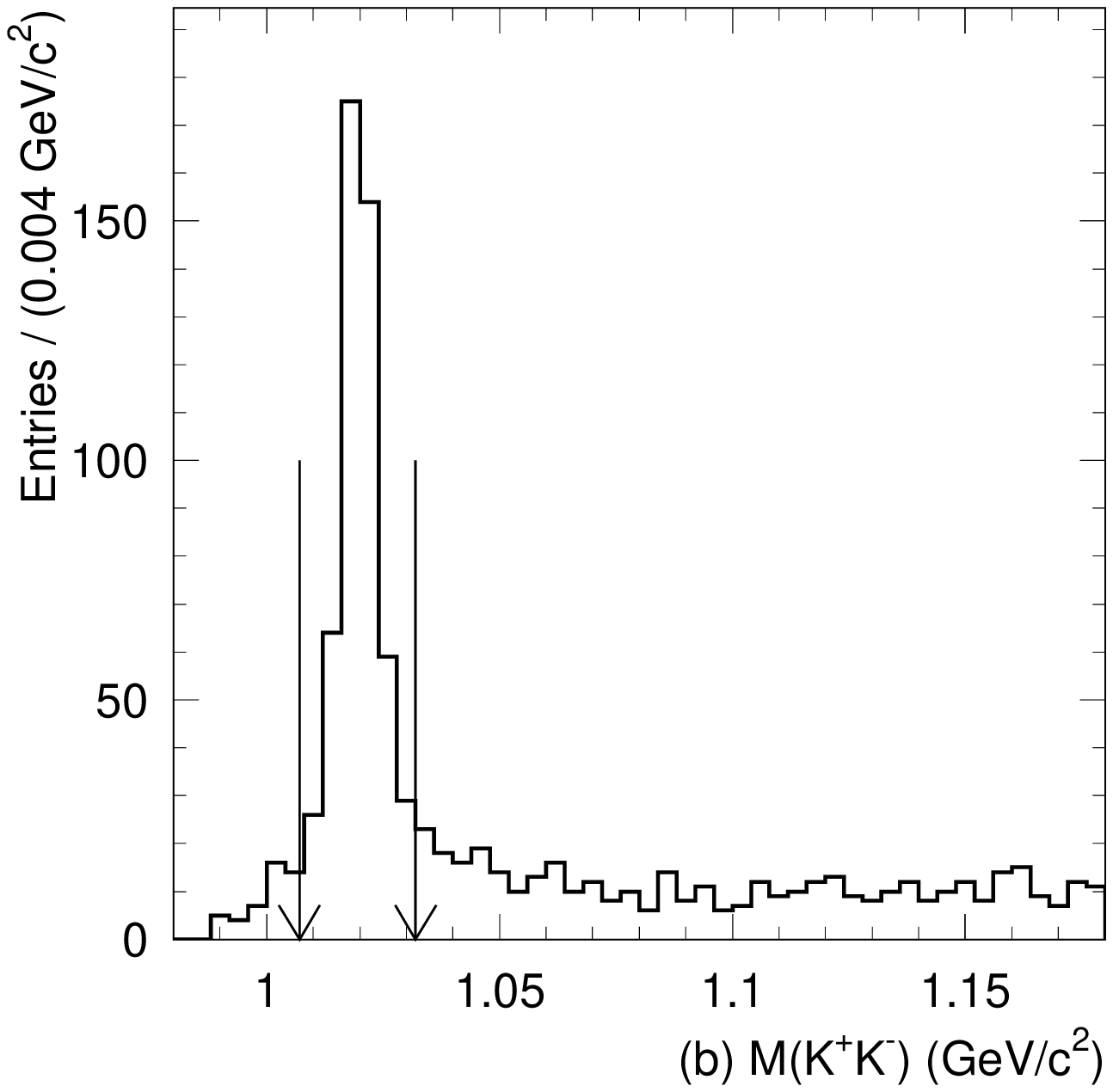,width= 7.5cm,height= 7.5cm,angle=0}
%   \psfig{file=test.eps,width= 7.5cm,height= 7.5cm,angle=0}
%                 \put(-40,100){(b)}
      \psfig{file=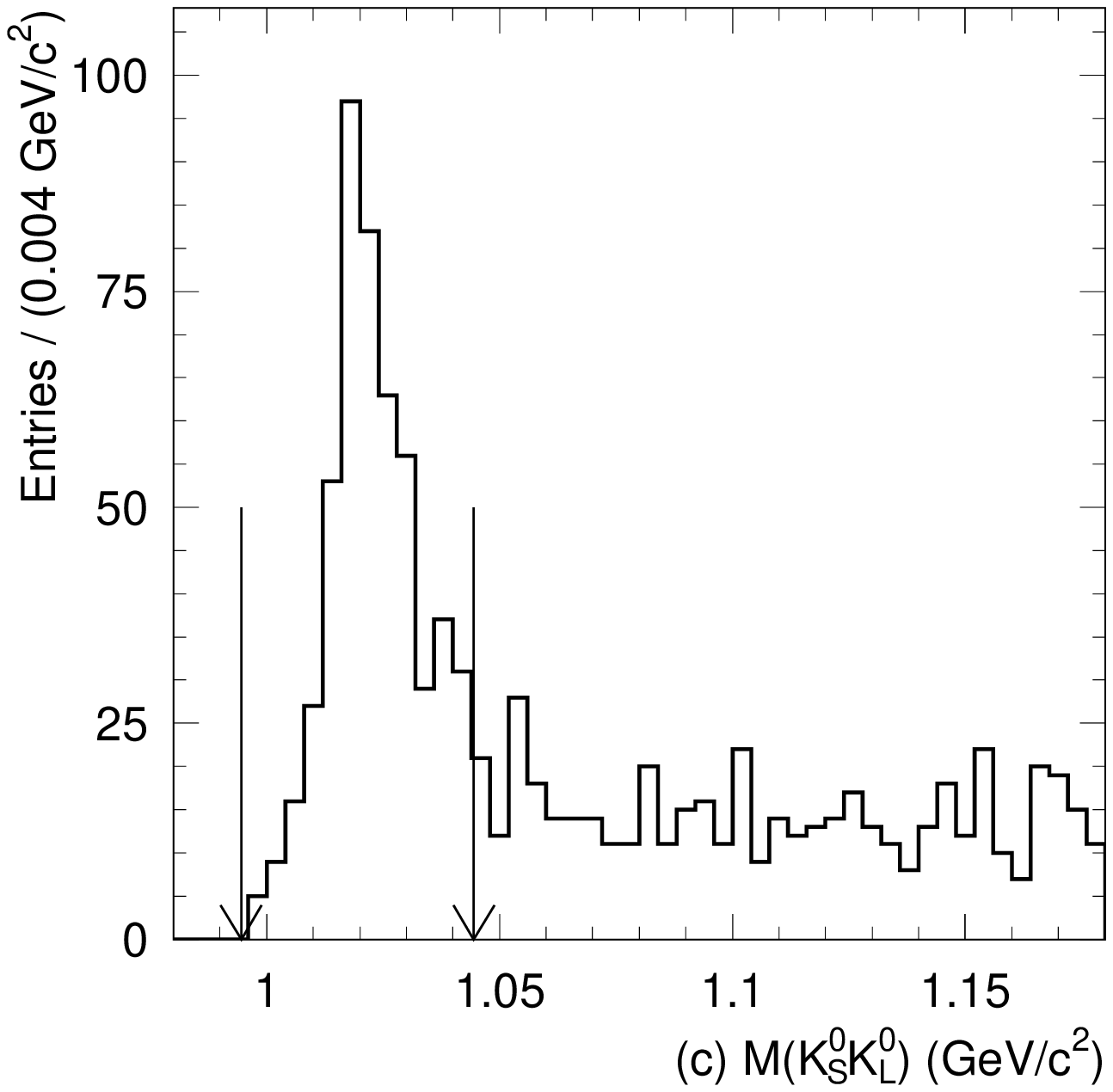,width= 7.5cm,height= 7.5cm,angle=0}
  %                  \put(-40,100){(c)}
  }
 %  \vskip -0.5cm
   \caption{
      (a) The $K^0_SK^0_L$ versus $K^+K^-$ invariant mass
   distribution.
   (b) The $K^+K^-$ invariant mass distribution when the $K^0_SK^0_L$
   invariant mass is in the $\phi\to K^0_SK^0_L$ signal region. The arrows show the $\phi\to K^+K^-$ signal region.
   (c) The $K^0_SK^0_L$ invariant mass distribution when the $K^+K^-$
   invariant mass is in the $\phi\to K^+K^-$ signal region. The arrows show the $\phi\to K^0_SK^0_L$ signal region.
}
   \label{mphi_sig}
\end{figure}

\begin{figure}[htbp]
 %  \vskip -1cm
   \centerline{
   \psfig{file=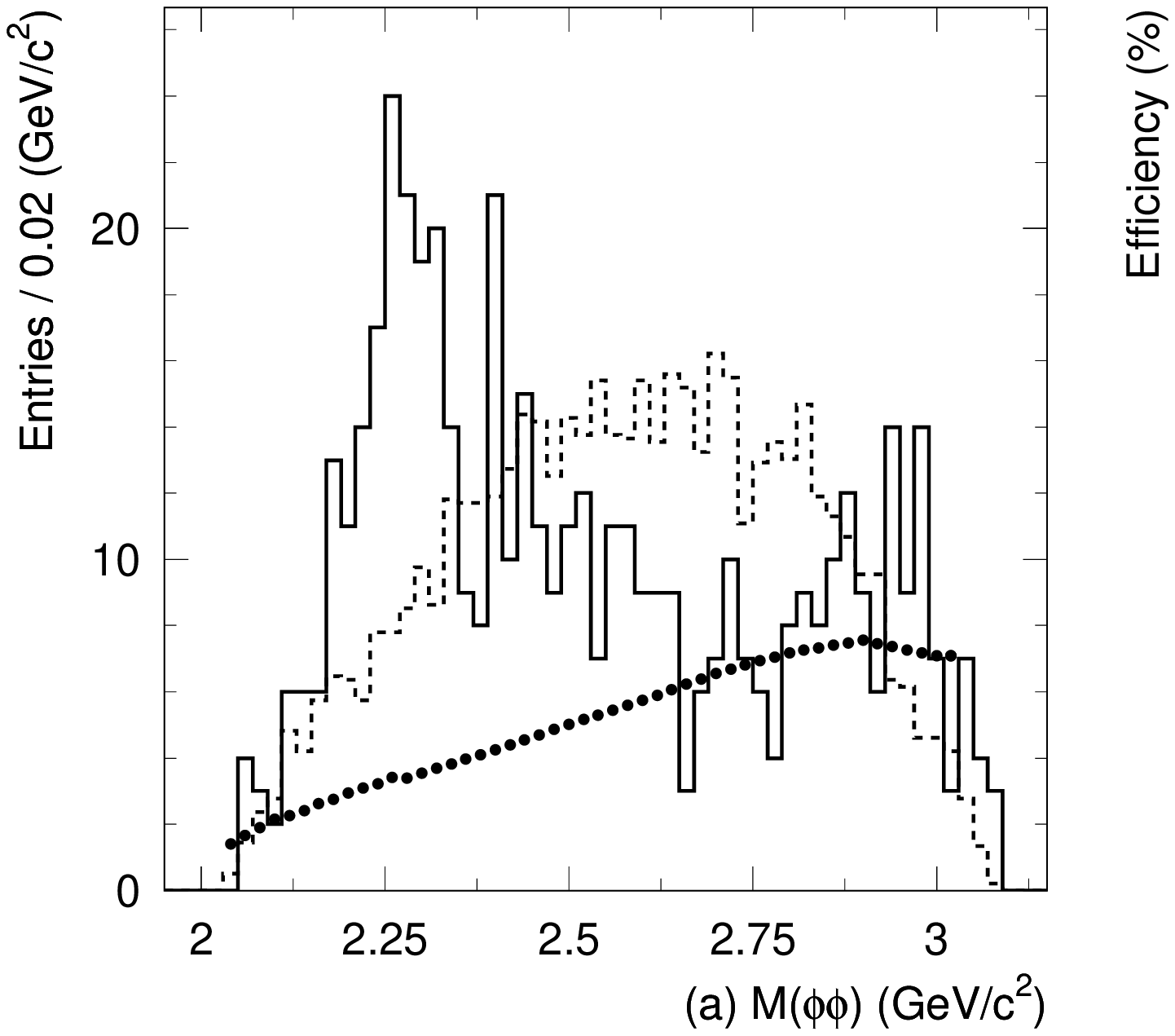,width=8.5cm,height=7.5cm,angle=0}
                 %\put(-50,175){(a)}
   \psfig{file=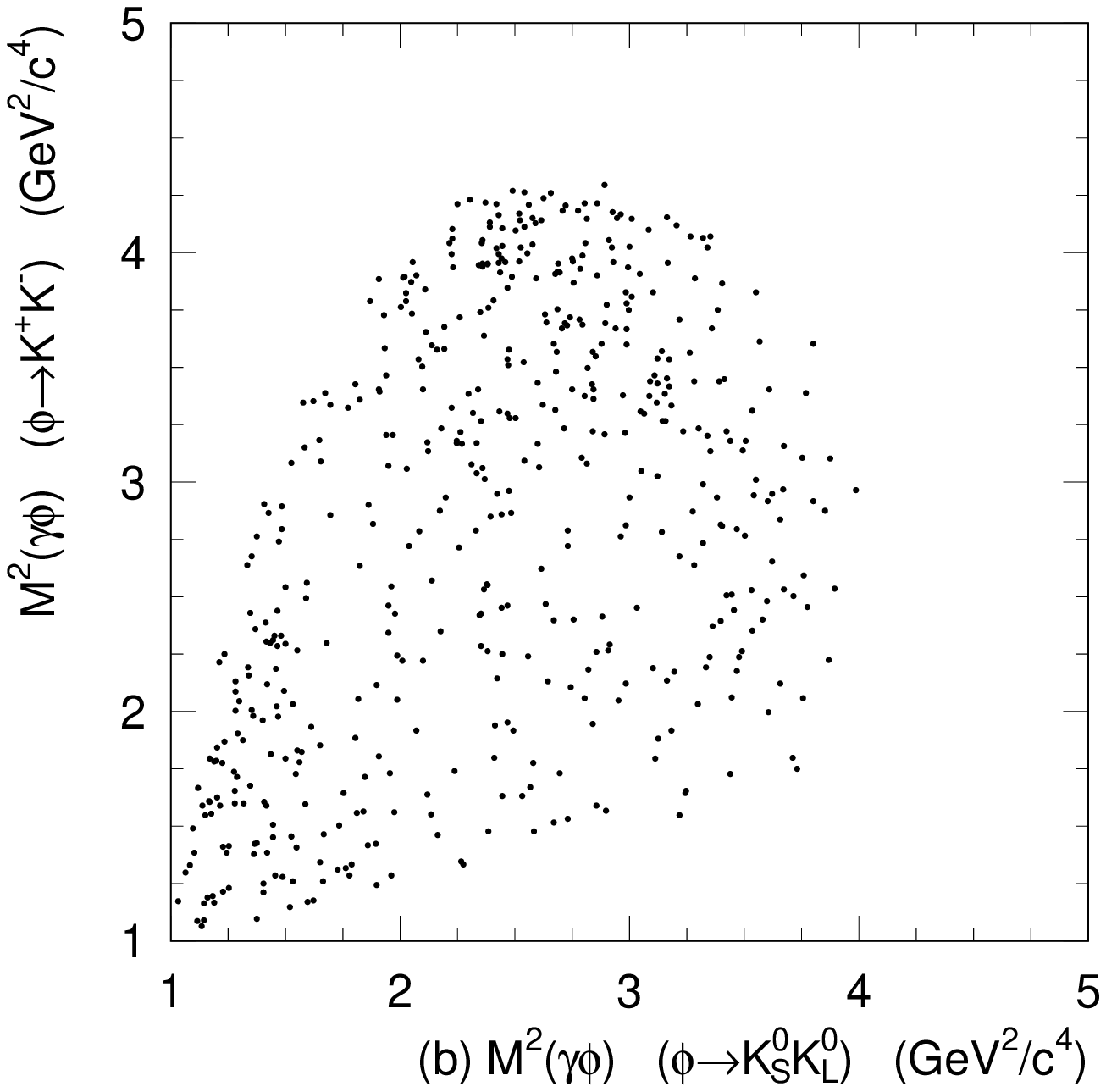,width=7.5cm,height=7.5cm,angle=0}
                 %\put(-50,175){(b)}
                 }
  % \vskip -0.5cm
   \caption{(a) The $K^+K^-K^0_SK^0_L$ invariant mass distribution for $J/\psi \to \gamma \phi\phi$ candidate events.
     The dashed histogram is the phase space invariant mass
     distribution, and
     the dotted curve indicates how the acceptance varies with the
     $\phi\phi$ invariant mass.  (b) Dalitz plot for $J/\psi \to
     \gamma \phi\phi$ candidate events. }
   \label{dalitz}
\end{figure}

The $K^0_SK^0_L$ versus $K^+K^-$ invariant masses are plotted in
Fig.~\ref{mphi_sig} (a), where a cluster of events corresponding to
$\phi\phi$ production is evident.  Because the processes $J/\psi \to
\phi\phi$ and $J/\psi \to \pi^0 \phi\phi$ are forbidden by
C-invariance, the presence of two $\phi$'s is a clear signal for the
radiative decay $J/\psi \to \gamma \phi\phi$. The histogram in
Fig.~\ref{mphi_sig} (b) shows the $K^+K^-$ invariant mass
distribution after the requirement that the $K^0_SK^0_L$ invariant
mass is inside the $\phi\to K^0_SK^0_L$ signal region
($|M(K^0_SK^0_L)-m_{\phi}|<0.025$~GeV/$c^2$). The histogram in
Fig.~\ref{mphi_sig} (c) shows the $K^0_SK^0_L$ invariant mass
distribution after the requirement that the $K^+K^-$ invariant mass
is inside the $\phi\to K^+K^-$ signal region
($|M(K^+K^-)-m_{\phi}|<0.0125$~GeV/$c^2$). The histogram in
Fig.~\ref{dalitz} (a) shows the $K^+K^-K^0_SK^0_L$ invariant mass
distribution for events where the $K^+K^-$ and $K^0_SK^0_L$
invariant masses lie within the $\phi\phi$ mass region
($|M(K^+K^-)-m_{\phi}|<0.0125$ GeV/$c^2$ and
$|M(K^0_SK^0_L)-m_{\phi}|<0.025$ GeV/$c^2$). There are a total of
508 events, which survive the above-listed criteria (optimized for
low $\phi\phi$ masses), with a prominent structure around
2.24~GeV/$c^2$. The phase space invariant mass distribution and the
acceptance versus $\phi\phi$ invariant mass are also shown in
Fig.~\ref{dalitz} (a) as the dashed histogram and dotted curve,
respectively. The peak is also evident as a diagonal band along the
upper right-hand edge of the Dalitz plot, shown in Fig.~\ref{dalitz}
(b). The asymmetry in the dalitz plot of data is caused by detection
efficiency.

\begin{figure}[htbp]
%   \vskip -1cm
   \centerline{
   \psfig{file=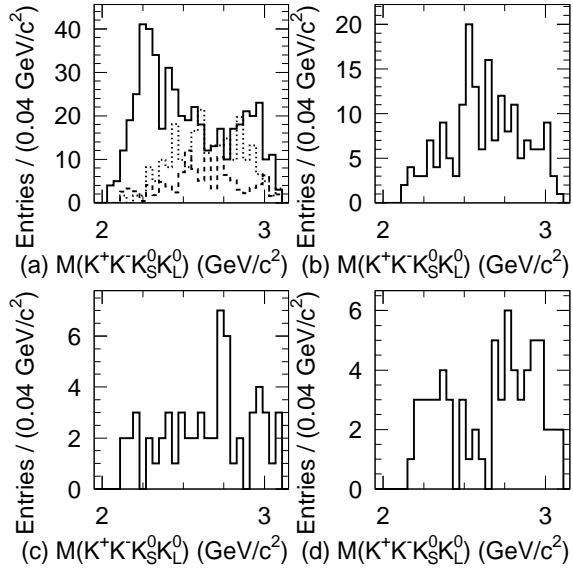,width=7.5cm,height=7.5cm,angle=0}}
%   \vskip -0.5cm
   \caption{The $K^+K^-K^0_SK^0_L$ invariant mass distributions for
(a) events in the $\phi\phi$ signal region; (b) events in the $\phi\to
K^+K^-$ sideband region; (c) events in the $\phi\to K^0_SK^0_L$
sideband region; (d) events in the corner region, as described in
the text. The dashed histogram in (a) shows the background
distribution obtained from the sideband evaluation. The dotted
histogram in (a) shows the $\phi\phi$ invariant mass distribution of
inclusive $J/\psi$ Monte Carlo samples }
   \label{m2phi_4}
\end{figure}

Non-$\phi\phi$ backgrounds are studied using events in the $\phi$
sideband regions shown in Fig.~\ref{mphi_sig} (a).
Figure~\ref{m2phi_4} (b) shows the $K^+K^-K^0_SK^0_L$ invariant mass
of events within the $\phi\to K^+K^-$ sideband region (1.069
GeV/$c^2$ $<M(K^+K^-)<1.094$ GeV/$c^2$ and
$|M(K^0_SK^0_L)-m_{\phi}|<0.025$ GeV/$c^2$), and Fig.~\ref{m2phi_4} (c)
shows the corresponding spectrum of events within the $\phi\to
K^0_SK^0_L$ sideband region ($|M(K^+K^-)-m_{\phi}|<0.0125$ GeV/$c^2$
and 1.082 GeV/$c^2$$<M(K^0_SK^0_L)<1.132$ GeV/$c^2$).
Figure~\ref{m2phi_4} (d) shows the events in the corner region, which
is defined as (1.069 GeV/$c^2$$<M(K^+K^-)<1.094$ GeV/$c^2$
 and
1.082 GeV/$c^2$$<M(K^0_SK^0_L)<1.132$ GeV/$c^2$). The background,
estimated by summing up the scaled event yields in Figs.
\ref{m2phi_4} (b)~and~(c) and subtracting that in Fig.
\ref{m2phi_4} (d),  is shown as the dashed histogram in Fig.
\ref{m2phi_4} (a). No sign of an enhancement near the $\phi\phi$ mass
threshold is evident in the non-$\phi\phi$ background events.

The backgrounds in the selected event sample are studied with Monte
Carlo simulations of a number of potential background decay channels
listed in the PDG tables~\cite{pdg2006}. The main background
originates from $J/\psi \to \phi K^{*}\overline{K}{}, K^{*} \to K
\pi^0$ (+ {\it
  c.c.}). Using the PDG's world average branching fraction for this
mode, we estimate that about 40 events from this channel are in the
$\phi \phi$ invariant mass signal region. However, the simulation
also shows that they do not peak at low $\phi\phi$ masses. Using a
Monte Carlo sample of 35.1M inclusive $J/\psi$ decay events
generated with the LUND-charm model ~\cite{lund}, we searched for
other possible background channels. None of the simulated channels
produce a peak at low $\phi\phi$ invariant masses. The dotted
histogram in Fig. \ref{m2phi_4} (a) shows the $\phi\phi$ invariant
mass distribution and the normalized background estimated with
inclusive $J/\psi$ Monte Carlo samples where the events of
$J/\psi\to\gamma\phi\phi$ and $J/\psi\to\gamma\eta_{c}
(\eta_{c}\to\phi\phi)$ are removed.

\section{\boldmath Partial wave analysis}
A partial wave analysis (PWA) of the events with $M(\phi\phi)<$
2.7~GeV/$c^2$ was performed. The two-body decay amplitudes in the
sequential decay process $J/\psi \to \gamma X, X\to \phi\phi$,
$\phi\to K^+ K^- $ and $\phi\to K^0_S K^0_L$ are constructed using
the covariant helicity coupling amplitude method~\cite{pwa}. The
intermediate resonance $X$ is described with the normal Breit-Wigner
propagator $\rm { BW } = 1/(\rm{ M }^2-s-i \rm{ M }\Gamma)$, where
$s$ is the $\phi\phi$ invariant mass-squared and $\rm M$ and
$\Gamma$ are the resonance's mass and width. The amplitude for the
sequential decay process is the product of all decay amplitudes and
the Breit-Wigner propagator. The total differential cross section
$d\sigma/d\Phi$ is

\begin{eqnarray}
\frac{d\sigma}{d\Phi} = |A(0^{-+})+A(0^{++})+A(2^{++})|^2 + BG,
\end{eqnarray}
 where $A(J^{PC})$ is the total amplitude for all resonances whose
spin-parity are $J^{PC}$,  and $BG$ denotes the background
contribution, which is described by a non-interfering phase space
term.
%The background is approximated by a non-interfering phase space term.

The relative magnitudes and phases of the amplitudes are determined
by an unbinned maximum likelihood fit.
%The details of the construction of
%the likelihood function are described in Ref.~\cite{guozj}, and
%MINUIT~\cite{minuit} is used to optimize the free parameters. In
%the minimization procedure, a change in log likelihood of 0.5
%represents a one standard deviation effect.
The basis of likelihood fitting is the calculation of the
probability that a hypothesized probability distribution function
would produce the data set under consideration. The probability to
observe the event characterized by the measurement $\xi$ is
\begin{eqnarray}
  P(\xi) = \frac{\omega(\xi)\epsilon(\xi)}
   {\int{d\xi}\omega(\xi)\epsilon(\xi)},
\end{eqnarray}
 where $\omega(\xi)\equiv\frac{d\sigma}{d\Phi}$ and $\epsilon(\xi)$
is the detection efficiency. The normalization integral
$\int{d\xi}\omega(\xi)\epsilon(\xi)$ is done with a weighted phase
space MC sample; the details are described in Ref.~\cite{guozj}. The
joint probability density for observing the $N$ events in the data
sample is
\begin{eqnarray}
  \mathcal{L} = \prod_{i=1}^{N}{P(\xi_i)}= \prod_{i=1}^{N}{\frac{\omega(\xi_i)\epsilon(\xi_i)}
   {\int{d\xi_i}\omega(\xi_i)\epsilon(\xi_i)}}.
\end{eqnarray}
For technical reasons, rather than maximizing $\mathcal{L}$,
$\mathcal{S} = -\rm{ln} \mathcal{L}$ is minimized, {\it i.e.}
\begin{eqnarray}
  \rm{ln}\mathcal{L} = \sum_{i=1}^{N}{\rm{ln}\displaystyle\bigg(\frac{\omega(\xi_i)}
   {\int{d\xi_i}\omega(\xi_i)\epsilon(\xi_i)}\displaystyle\bigg)} + \sum_{i=1}^{N}{\rm{ln}\epsilon(\xi_i)} .
\end{eqnarray}
%%Here you should give a more explicit represenation of P(x_1)
For a given data set, the second term is a constant and has no
impact on the determination of the parameters of the amplitudes or
on the relative changes of $\mathcal{S}$ values. So, for the
fitting, $\rm{ln}\mathcal{L}$, defined as:
\begin{eqnarray}
  \rm{ln}\mathcal{L} \equiv \sum_{i=1}^{N}{\rm{ln}\displaystyle\bigg(\frac{\omega(\xi_i)}
   {\int{d\xi_i}\omega(\xi_i)\epsilon(\xi_i)}\displaystyle\bigg)},
\end{eqnarray}
is used. The free parameters are optimized by MINUIT~\cite{minuit}.
In the minimization procedure, a change in log likelihood of 0.5
represents a one standard deviation effect for the one parameter case.

For the production of a pseudoscalar, only $\mathcal{P}$ waves are
allowed in both the radiative decay $J/\psi \to \gamma X$ and  the
hadronic decay $X\to \phi\phi$. For the case of a scalar, both
$\mathcal{S}$ and $\mathcal{D}$ waves are possible in both the radiative
and hadronic decays; only the $\mathcal{S}$ wave is considered in
the fit. For the production of a $2^+$ resonance, there are five
possible amplitudes for both the radiative and the hadronic decays,
one $\mathcal{S}$ wave, three $\mathcal{D}$ waves and one $(L=4)$
wave. In this case, only $\mathcal{S}$ and $\mathcal{D}$ waves in
both decays, corresponding to the lower overall spin of the
$\phi\phi$ system, are considered. (The $(L=4)$ wave is ignored in
this analysis.)

\begin{figure}[htbp]
%   \vskip -1cm
   \centerline{
   \psfig{file=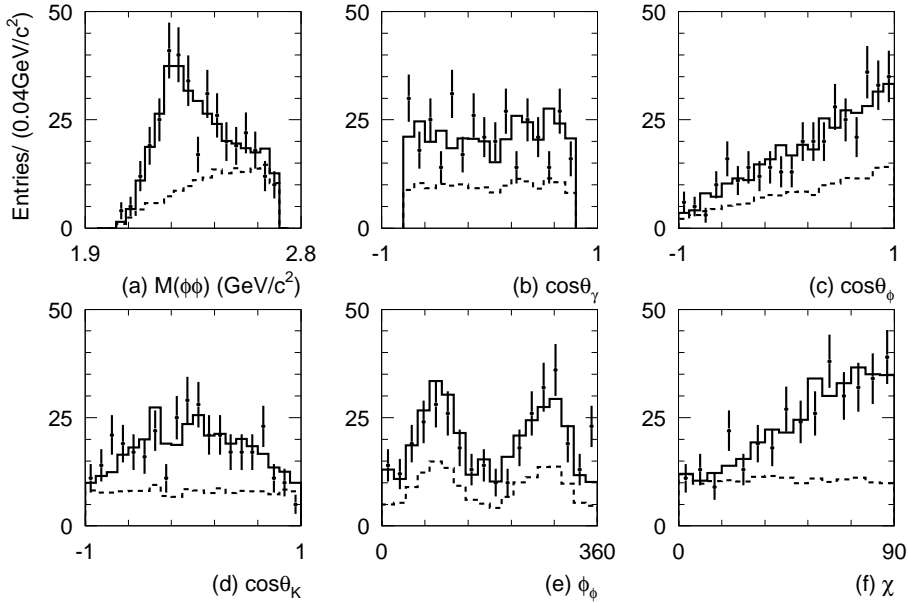,width=12cm,height=8cm,angle=0}
                % \put(-300,200){(a)}
                % \put(-190,200){(b)}
                % \put(-75,200){(c)}
                % \put(-300,90){(d)}
                % \put(-190,90){(e)}
                % \put(-75,90){(f)}
}
%   \vskip -0.5cm
   \caption{Comparisons between data and MC projections
for the $\phi\phi$ invariant mass distribution and the angular
distributions for events with $\phi\phi$ invariant mass less
than 2.7 GeV/$c^2$ using the fitted $\eta(2225)$ resonance parameters.
The points with error bars are data, the solid
histograms are the MC projections, and the dashed lines are the
background contributions. (a) The $\phi\phi$ invariant mass
distribution; (b) the polar angle of the radiative photon
($\theta_{\gamma}$); (c) the polar angle of the $\phi$ in the
$\phi\phi$ rest system ($\theta_{\phi}$); (d) the polar angle of the
kaon in the $\phi$ rest system ($\theta_K$); (e) the azimuthal angle
 of the $\phi$ in the $\phi\phi$ rest system ($\phi_\phi$); and (f) the $\chi$
distribution, where $\chi$ is the azimuthal angle between the normal
directions of the two decay planes in the $\phi\phi$ rest frame.}
   \label{proj}
\end{figure}

When $J/\psi \to \gamma X$, $X \to \phi \phi$ is fitted with both
the $\phi\phi$ and $\gamma X$ systems in a $\mathcal{P}$ wave,
which corresponds to a $X = 0^{-+}$ pseudoscalar state,
the fit gives $195.5^{+18.6}_{-18.7}$ events with mass $M =
2.24^{+0.03}_{-0.02}$~GeV/$c^2$, width $\Gamma =
0.19 \pm 0.03$~GeV/$c^2$, and a statistical significance
larger than 10~$\sigma$.
%The mass and width of the resonance are obtained from the optimization.
The errors are statistical only.
% and the correlations between the different resonances are included.
Using a selection efficiency of 3.29\%, which is determined from the
Monte-Carlo simulation using the magnitudes and phases of the
partial amplitudes from the PWA, we obtain a product branching
fraction of:
$$ Br(J/\psi \to \gamma \eta(2225))\cdot Br(\eta(2225)\to
\phi\phi)=(4.4 \pm 0.4)\times 10^{-4}. $$ Details of the fitting
procedure and the detection efficiency determination can be found in
Ref.~\cite{guozj}.
%Comparisons of the
%$\phi\phi$ invariant mass and the angular distributions between
%data and Monte Carlo projections with fitted parameters have been
%performed.
Figure~\ref{proj} (a) shows a comparison of the data and MC
projections of the $\phi\phi$ invariant mass distribution for the
fitted parameters. Comparisons of the projected data and MC angular
distributions for the events with $\phi\phi$ invariant mass less
than 2.7 GeV/$c^2$ are shown in Figs.~\ref{proj} (b)-(f).

We also tried to fit the resonance with $0^{++}$ and $2^{++}$
spin-parity hypotheses using all possible combinations of orbital
angular momenta in the $\phi\phi$ and $\gamma (\phi\phi)$ systems.
The log-likelihood values of the best fits are worse than that of the
$0^{-+}$ assignment by 95 and 27 for the $0^{++}$ and $2^{++}$
assignments, respectively.  We therefore conclude that the $J^{PC}$
of the resonance strongly favors $0^{-+}$ ($\eta(2225)$).

Because of the possible existence of a coherent $J/\psi\to \gamma
\phi\phi$ phase space amplitude, we also fitted with an interfering
phase space ($0^-$) term included. The log-likelihood value
$\mathcal{S}$ improves by 0.4, which suggests that the contribution
from $0^{-+}$ phase space is negligible.

If an additional resonance is included in the fit, the significance of
the additional resonance after reoptimization is 0.8 $\sigma$, 2.1
$\sigma$ and 3.3 $\sigma$ for the $0^{-+}$, $0^{++}$ and $2^{++}$
assignments of the additional resonance, respectively. The differences
between the results including and not including the additional
$0^{++}$ or $2^{++}$ are included in the systematic errors; the
systematic uncertainty contributions for the mass, width, and
branching fraction of the $\eta(2225)$ are ${+0.01}$~GeV/$c^{2}$,
${+0.04}$~GeV/$c^{2}$, and ${+2.6}\%$, respectively.

\section{ \boldmath Systematic error}
The systematic uncertainties are estimated by considering the
following: the uncertainties in the modeling of the background,
different Breit-Wigner parameterizations, the possible presence of
additional resonances, the simulation of the MDC wire resolution, as
well as possible biases in the fitting procedure. The uncertainties
in the background include the uncertainty in the treatment of the
background in the fitting. We also tried to subtract the backgrounds
determined from the sidebands in the fit, and the differences are
taken as systematic errors. Since the enhancement is in the
near-threshold region of the $\phi\phi$ invariant mass spectrum, a fit
using a Breit-Wigner with a momentum dependent width~\cite{sbw} is
also performed, and the differences between the fit with a constant
width Breit-Wigner are included as systematic errors.
Possible fitting biases are estimated from the differences obtained between
input and output masses and widths from Monte Carlo samples, which
are generated as $J/\psi \to \gamma 0^{-+}, 0^{-+} \to \phi\phi$
using the fitted parameters. The total systematic errors are obtained by adding
the individual errors in quadrature. The total systematic errors on
the mass and width are determined to be
$^{+0.03}_{-0.02}$~GeV/$c^{2}$ and $^{+0.06}_{-0.04}$~GeV/$c^{2}$,
respectively.

For the systematic error on the branching fraction measurement, the
systematic uncertainties of the photon detection efficiency and the
particle identification efficiency, as well as the $\phi$ and
$K^0_S$ decay branching fractions, the mass and width uncertainties
of $\eta(2225)$, and the total number of $J/\psi$
events~\cite{jpsinum} are also included. The total relative
systematic error on the product branching fraction is
$^{+17.7}_{-18.5}\%$.

\section{ \boldmath Summary}
Using $5.8 \times 10^7 J/\psi$ events measured in the BESII
detector, we  studied the radiative decay $J/\psi \to \gamma \phi
\phi \to \gamma K^+ K^- K^0_S K^0_L$. A structure
($\eta(2225)$) in the near-threshold region of $\phi\phi$ invariant
mass spectrum is observed.

A PWA shows that the structure is dominated by a $0^{-+}$ state with
a mass $2.24^{+0.03}_{-0.02}{}^{+0.03}_{-0.02}$ GeV/$c^{2}$ and a
width $0.19\pm0.03^{+0.06}_{-0.04}$~GeV/$c^{2}$. The product
branching fraction is measured to be: $$Br(J/\psi \to \gamma
\eta(2225))\cdot Br(\eta(2225)\to \phi\phi)
 = (4.4\pm0.4\pm0.8)\times 10^{-4}.$$

\section{ \boldmath Acknowledgments}
The BES collaboration thanks the staff of BEPC and computing center
for their hard efforts. This work is supported in part by the
National Natural Science Foundation of China under contracts Nos.
10491300, 10225524, 10225525, 10425523, 10625524, 10521003, the
Chinese Academy of Sciences under contract No. KJ 95T-03, the 100
Talents Program of CAS under Contract Nos. U-11, U-24, U-25, and the
Knowledge Innovation Project of CAS under Contract Nos. U-602, U-34
(IHEP), the National Natural Science Foundation of China under
Contract No. 10225522 (Tsinghua University), and the Department of
Energy under Contract No. DE-FG02-04ER41291 (U. Hawaii).

\end{document}